\documentclass{article}

     \usepackage[preprint,nonatbib]{neurips_2023}

\usepackage[utf8]{inputenc} 
\usepackage[T1]{fontenc}    
\usepackage{hyperref}       
\usepackage{url}            
\usepackage{booktabs}       
\usepackage{amsfonts}       
\usepackage{nicefrac}       
\usepackage{microtype}      
\usepackage{xcolor}         
\usepackage{caption}        
\usepackage{subcaption}     
\usepackage{graphicx}       
\usepackage{wrapfig}        
\usepackage[backend=biber, style=ieee]{biblatex} 
\usepackage{xparse,mathtools}
\title{Unsupervised segmentation of irradiation-induced order–disorder phase transitions in electron microscopy}

\author{
  Arman H Ter-Petrosyan \\
  Pacific Northwest National Laboratory\\
  Richland WA 99354 \\
  \texttt{arman.ter-petrosyan@pnnl.gov} \\
  \And 
  Jenna A Bilbrey \\
  Pacific Northwest National Laboratory\\
  Richland WA 99354 \\
  \texttt{jenna.pope@pnnl.gov} \\
  \And 
  Christina M Doty \\
  Pacific Northwest National Laboratory\\
  Richland WA 99354 \\
  \texttt{christina.doty@pnnl.gov} \\
  \And 
  Bethany E Matthews \\
  Pacific Northwest National Laboratory\\
  Richland WA 99354 \\
  \texttt{bethany.matthews@pnnl.gov} \\
  \And
  Le Wang \\
  Pacific Northwest National Laboratory\\
  Richland WA 99354 \\
  \texttt{le.wang@pnnl.gov} \\
  \And 
  Yingge Du \\
  Pacific Northwest National Laboratory\\
  Richland WA 99354 \\
  \texttt{yingge.du@pnnl.gov} \\
  \And   
  Eric Lang \\
  University of New Mexico\\
  Albuquerque NM 87131\\
  \texttt{ejlang2@unm.edu} \\
  \And   
  Khalid Hattar \\
  University of Tennessee - Knoxville\\
  Knoxville TN 37996 \\
  \texttt{khattar@utk.edu} \\
  \And 
  Steven R Spurgeon \\
  Pacific Northwest National Laboratory\\
  Richland WA 99354 \\
  \texttt{steven.spurgeon@pnnl.gov} \\
}

\begin{document}

\maketitle{}

\begin{abstract}
  We present a method for the unsupervised segmentation of electron microscopy images, which are powerful descriptors of materials and chemical systems. Images are oversegmented into overlapping chips, and similarity graphs are generated from embeddings extracted from a domain-pretrained convolutional neural network (CNN). The Louvain method for community detection is then applied to perform segmentation. The graph representation provides an intuitive way of presenting the relationship between chips and communities. We demonstrate our method to track irradiation-induced amorphous fronts in thin films used for catalysis and electronics. This method has potential for ``on-the-fly'' segmentation to guide emerging automated electron microscopes.
\end{abstract}


\section{Introduction}

The development of new technologies for energy storage, quantum computing, and many other applications is predicated on the design of better performing materials. It is important to understand not just the intrinsic structure of materials, but also how they evolve in real-world environments of high temperature, radiation, and pressure \cite{eswarappa2023materials}. Scanning transmission electron microscopy (STEM) is a data-rich analytic technique to study the dynamic evolution of materials down to the atomic level.\cite{spurgeon2020order} Recent interest in coupling STEM with machine learning (ML)-driven analytics to process and interpret microscopy images in real time has opened a path toward autonomous \textit{in situ} STEM experimentation \cite{ziatdinov2021atomai, liu2022experimental, spurgeon2021towards, olszta2022automated, zheng2022situ, cheng2022review}. Because it is difficult to procure high-quality labeled training data, many past approaches have relied on synthetic data,\cite{williamson2022creating} while some have focused on sparse data approaches \cite{chen2023small, akers2021rapid}. It is becoming increasingly clear that unsupervised approaches that do not rely on extensive \textit{a priori} knowledge of the sample are key to creating a general analysis framework for microscopy data \cite{shi2022uncovering, williamson2022creating}.

ML has shown promise for the common workflow of identifying structural defects in STEM images with high throughput and veracity \cite{dan2019machine}, facilitating the discovery of structure-property relationships in materials systems \cite{kalinin2022deep}. Complex oxides are compelling candidates for such analysis, as their valuable technological properties are sensitive to even trace amounts of defects \cite{gunkel2020oxygen, tuller2011point}. When these materials are deployed, radiation can increase defect populations \cite{sickafus2000radiation}, impacting performance and leading to early device failure. To this end, the various structural regions present in the irradiated material must first be quantified, a task well suited to data-driven STEM analysis. However, we presently lack robust ML analytics to generalizably describe such defect formation, which is necessary to derive more accurate physical models and improve material performance.

Here, we present a method for the unsupervised segmentation of structural regions within STEM images through the combination of pretrained classification models and graph theory. Embeddings of overlapping superpixels are extracted from a ResNet50 model pretrained on a large dataset of microscopy images \cite{stuckner2022microstructure}. A graph is formed based on the pairwise similarity of the embeddings. The Louvain method \cite{blondel2008fast} is used to identify communities within the graph, which correspond to various structural regions in the material. This approach provides a means to analyze electron microscopy data absent \textit{a priori} knowledge and shows promise to describe order-disorder phase transitions in emerging functional materials.


\section{Related Work}

Supervised segmentation of crystalline materials has been widely performed for various types of electron microscopy images \cite{li2021defectnet, stuckner2022microstructure, ziatdinov2021atomai, sadre2021deep, groschner2021machine, lee2022stem}, while unsupervised segmentation has been comparatively less studied. Akers et al.~developed pyCHIP, a few-shot approach for oversegmentation of STEM images that is robust against noise and requires less data than conventional segmentation methods \cite{akers2021rapid}. In the pyCHIP method, an image is oversegmented into non-overlapping squares, embeddings are extracted from a pretrained classification model and the embeddings of each chip are compared against a small set of hand-selected prototype chips. We note that the unsupervised method discussed in this paper moves beyond pyCHIP, since the present approach does not require labeled data and utilizes a microscopy-specific encoder to compute embeddings. Other unsupervised approaches for STEM images have applied statistical clustering methods to some type of descriptor. For instance, Wang et al.~applied principal component analysis (PCA) and k-means clustering on local features obtained from symmetry operations for unsupervised clustering of STEM high-angle annular dark field (STEM-HAADF) images \cite{wang2021segmentation}. To identify nanoparticles in STEM images, Wang et al.~applied k-means clustering followed by naive Bayes classification to a set of geometric shape descriptors \cite{wang2021autodetect}.

It has been recently shown that the unsupervised segmentation of digital photographs can be achieved by combining image features or learned embeddings and graph theory. Jiao et al.~applied graph clustering to high-dimensional embeddings obtained from an autoencoder fed low-level visual features of color, edge, gradient, and image saliency \cite{jiao2020segmentation}.
Melas-Kyriazi et al.~performed spectral clustering on a graph Laplacian matrix formed from a combination of color information and unsupervised deep features \cite{melas2022deep}. Aflalo et al.~incorporated node features derived from pre-trained neural networks into the graph and performed clustering using a graph neural network trained on a loss function based on classical graph clustering algorithms \cite{aflalo2022deepcut}. While graph-based image segmentation existed prior to the widespread use of deep-learning-based segmentation \cite{felzenszwalb2004efficient, camilus2012review}, these approaches have not yet been widely applied to electron microscopy data.

\section{Methods}

\begin{figure}[hb]
    \centering
    \includegraphics[width=0.85\textwidth]{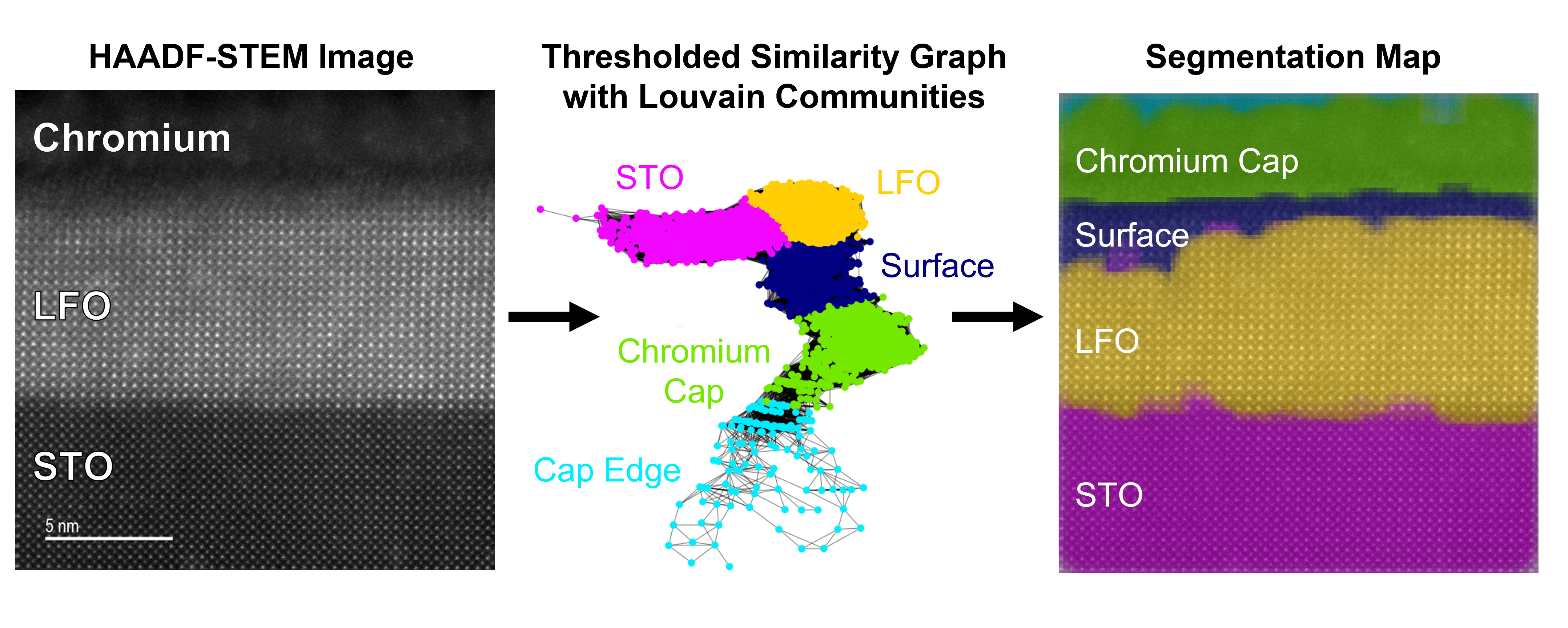}
    \caption{Workflow of the unsupervised segmentation method. A STEM-HAADF image is decomposed into overlapping chips, which represent a node in the similarity graph. Edges are determined based on the similarity between the embeddings of chip pairs. Louvain community detection is applied to the graph, and the segmentation map is generated by overlaying the community of each chip onto the STEM-HAADF image. The STEM-HAADF image is taken along the STO $\begin{bmatrix} 1 & 0 & 0 \end{bmatrix}$ zone-axis.}
    \label{fig:graphexample}
\end{figure}

Figure \ref{fig:graphexample} demonstrates the workflow involved in our unsupervised segmentation method. First, a STEM-HAADF image is oversegmented into overlapping square chips containing a $4\times4$ region of the atomic lattice. Embeddings for each chip are extracted from a CNN with a ResNet50 architecture pretrained on the MicroNet dataset \cite{stuckner2022microstructure}. Pairwise distances are computed in the embedding space for each chip using the cosine distance metric. The distances are normalized into similarity scores on the unit interval $[0, 1]$. A score of ``$0$'' corresponds to the most dissimilar pair of chips whereas a score of ``$1$'' corresponds to the most similar pair of chips in the image. The scores can naturally be organized using a complete graph in which all chips are represented as nodes and edges are weighted by the similarity scores. The edges can be further filtered to reduce the connectivity of the graph and produce communities of similar chips by removing edges with weights below a certain threshold value. Formatting the chip embedding similarities as a graph enables the use of graph-based clustering methods to identify regions in the image. Here, we use the Louvain method, which aims to optimize the modularity, which provides a measure of the division of the graph into discrete communities.



\section{Results \& Discussion}

To demonstrate this method, we examine STEM-HAADF images of an epitaxial thin film at three levels of irradiation---0, 0.1, and 0.5 displacements per atom (dpa)---which introduces increasing structural disorder. The single-crystal substrate consists of SrTiO$_3$ (STO), on which a perovskite LaFeO$_3$ (LFO) thin film was grown \cite{taylor2023resolving}. The material was then irradiated to 0.1 and 0.5 dpa, after which a protective Cr cap was sputtered on the film surface. Cross-sectional STEM samples were then prepared, as described in \cite{taylor2023resolving}. The Cr cap is polycrystalline and does not display a lattice structure in the STEM-HAADF image, in contrast to the LFO and STO layers which show clear perovskite lattices prior to irradiation.

\begin{figure}[ht]
    \centering
    \includegraphics[width=\textwidth]{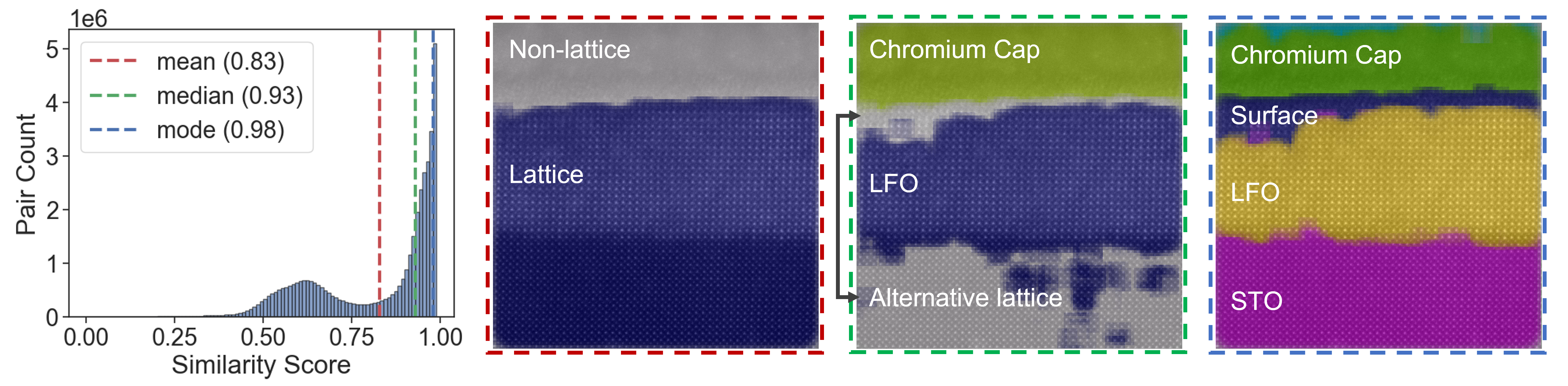}
    \caption{Similarity distribution and segmentation maps for the film prior to irradiation (0 dpa) thresholded at the mean (red), median (green), and mode (blue) of the similarity distribution.}
    \label{fig:thresholding}
\end{figure}

To reduce connectivity and expose partitions of similar chips, thresholding is applied to the complete graph prior to using the Louvain method. The choice of thresholding value has a noticeable impact on the communities detected by the Louvain method. Because the similarity distribution will be different for each image, we examine a series of thresholding values relative to the distribution: the mean, median, and mode. As demonstrated in Figure \ref{fig:thresholding} of the image prior to irradiation, increasing the thresholding value results in more fine-grained segmentation of the image into the layers visible in the original STEM-HAADF image.

Notably, thresholding at the lowest level (the mean) segments the image into two regions: one representing material with a clear crystalline lattice and one representing the chromium cap where no lattice is visible. Increasing the thresholding value from the mean to the median of the similarity distribution leads to segmentation of LFO in the crystalline region. Finally, when increasing the thresholding value from the median to the mode, both the LFO and STO layers as well as the surface are clearly differentiated.

\begin{wrapfigure}[22]{r}{0.6\textwidth}
    \centering
    \vspace{-4pt}
    \includegraphics[width=0.6\textwidth]{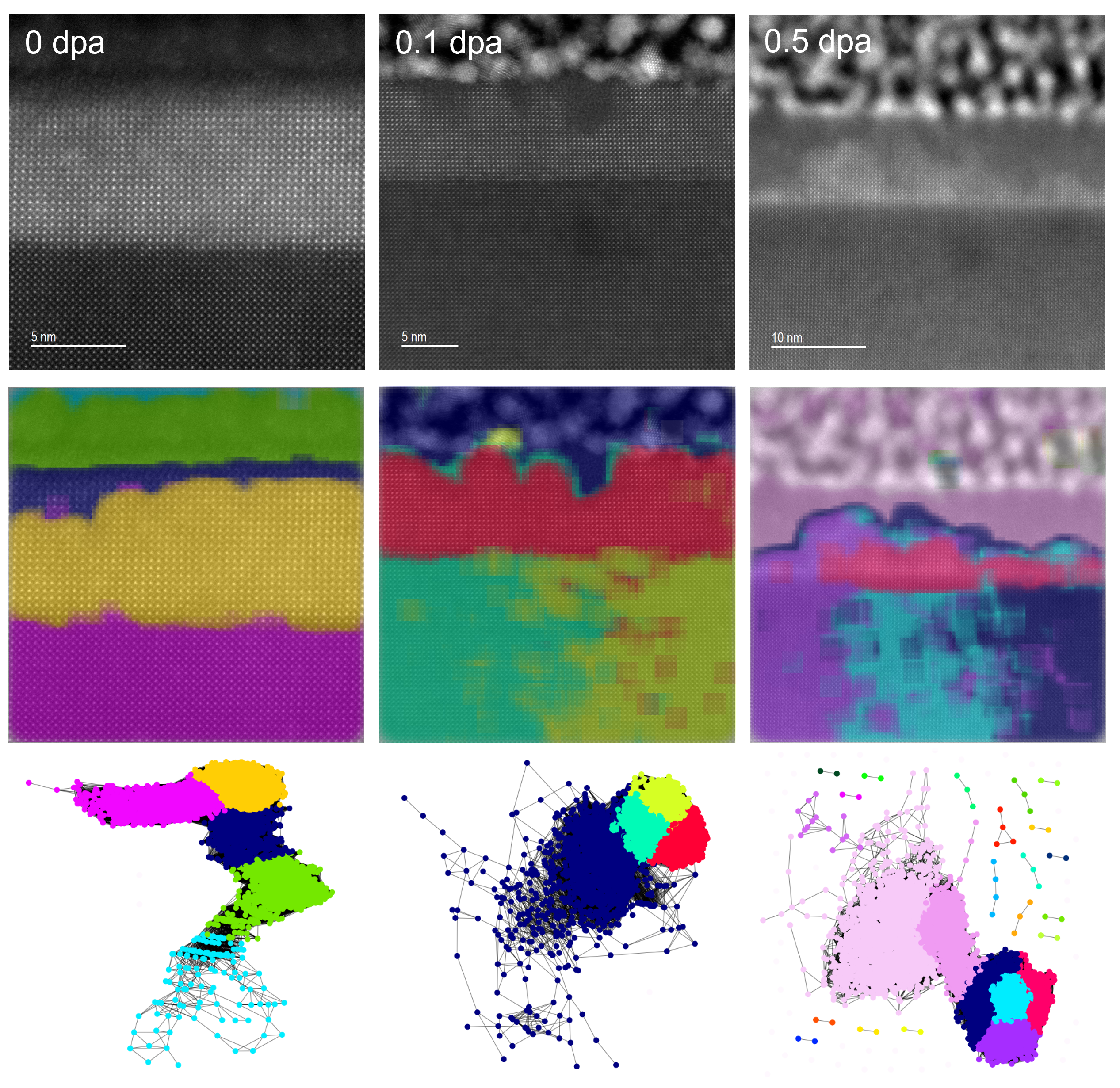}
    \vspace{-14pt}
    \caption{Cross-sectional STEM-HAADF images (top), segmentation maps (middle), and similarity graphs (bottom) for films irradiated to 0 (left), 0.1 (center), and 0.5 dpa (right).}
    \label{fig:LFOpristineirrad}
\end{wrapfigure}

Irradiation of the surface induces atomic rearrangement with associated loss of crystallinity (amorphization) \cite{spurgeon2020order}, creating an amorphous front that travels further down the lattice with increased irradiation. Figure \ref{fig:LFOpristineirrad} shows the segmentation results for samples irradiated at 0, 0.1, and 0.5 dpa. Edge thresholding was applied at the mode of the corresponding similarity distribution. The segmentation shows a demarcation between the amorphous and crystalline regions. Prior to irradiation, this demarcation relates to the interface between the LFO and the chromium cap. After irradiation, this demarcation follows the line of amorphization in the LFO layer. Amorphous regions in the LFO layer appear as sections of near uniform intensity, indicating a lack of ordered atomic columns. Varying levels of disorder produce images somewhere between the highly ordered atomic columns and completely amorphized regions. Our segmentation method identifies varying levels of amorphization in the LFO layer of samples irradiated to 0.1 and 0.5 dpa.

In the irradiated samples, STO does not segment into a single region. We believe this may be an artifact of a loss of focus in the lateral scan direction. Prior to irradiation, the focus is consistent across the image. For the 0.1 dpa sample, the right side of the image is more in focus than the left, and for the 0.5 dpa sample, the left side is more in focus than the right. These minor differences in focus results in slightly reduced similarity values. Notably, the communities comprising the STO layer are tightly integrated in the similarity graphs of the irradiated samples. Therefore, the identification of separate communities in the STO phase may be an artifact of the use of a heuristic algorithm for modularity optimization by the Louvain method.

\vskip 0.3in

\section{Conclusions}
In this work, we present a method for the unsupervised segmentation of STEM-HAADF images. Images are oversegmented into overlapping chips, and embeddings for each chip are extracted from a CNN pretrained on a microscopy dataset. A graph is generated based on the similarity of embeddings, to which edge thresholding and the Louvain method are applied. The identified communities represent segmented regions in the image. Notably, we show that our method can be applied to identify irradiation-induced amorphization in atomic-resolution STEM images. 

Because the resulting similarity scores depend on the encoder from which the chip embeddings are extracted, the use of different encoders will lead to variations in the similarity graph, which could either improve or weaken the segmentation results. In our case, the encoder trained on a microscopy-specific dataset produced effective microstructure feature representations that facilitated the identification of amorphous fronts in STEM images. 
As noted in prior work on the segmentation of microscopy data \cite{akers2021rapid}, this approach is sensitive to microscope parameters \cite{sytwu2023generalization}, as well as carbon contamination \cite{lee2022stem}, which must be carefully considered to obtain reliable results. Furthermore, the incorporation of additional modalities has potential to improve classifier performance. For instance, energy-dispersive X-ray spectroscopy (EDS), which identifies elemental species, will aid in the separation of both discrete material layers and amorphized regions subject to radiation-induced segregation.

\section{Broader Impact}
Segmentation is a critical step in the determination of structure-property relationships in many important materials and chemical systems. Often, this segmentation is performed by domain experts, limiting throughput and standardization efforts. Present applications of ML in materials characterization problems have typically relied on training neural networks for specific tasks, thus lacking transferability to other tasks. In many instances, there is not enough data to produce reliable training results. Moreover, when there is enough data to train models, it must be painstakingly labeled by a domain expert. We aim to ultimately eliminate the need for a user to prepare labeled data, even small amounts like what is required by few-shot approaches. Thus, transitioning from a semi-supervised, few-shot approach to a completely unsupervised, clustering approach to microscopy image segmentation. Though the current approach is relatively slow compared to the expert-labeled few-shot approach, further workflow optimization can help facilitate the use of this method in an automated STEM platform \cite{olszta2022automated, creange2022towards}.

\section*{Author Contributions}
A.H.T-P.~contributed to method development, method application, and writing of the manuscript.
J.A.B.~contributed to method development and writing of the manuscript.
C.M.D.~performed data preprocessing.
B.E.M.~collected the STEM images.
L.W.~and Y.D.~fabricated the LFO-STO films.
E.L.~and K.H.~irradiated the films.
S.R.S.~oversaw the work and contributed to method development and writing of the manuscript.

\begin{ack}
This research was supported by the Chemical Dynamics Initiative, under the Laboratory Directed Research and Development (LDRD) Program at Pacific Northwest National Laboratory (PNNL). PNNL is a multi-program national laboratory operated for the U.S. Department of Energy (DOE) by Battelle Memorial Institute under Contract No. DE-AC05-76RL01830. The growth of thin film samples was originally supported by DOE Office of Science, Basic Energy Sciences under award \#10122. STEM sample preparation was performed at the Environmental Molecular Sciences Laboratory (EMSL), a national scientific user facility sponsored by the Department of Energy's Office of Biological and Environmental Research and located at PNNL. STEM data was collected in the Radiological Microscopy Suite (RMS), located in the Radiochemical Processing Laboratory (RPL) at PNNL. Ion irradiation work was performed at the Center for Integrated Nanotechnologies, an Office of Science User Facility operated for the U.S. DOE. Sandia National Laboratories is a multimission laboratory managed and operated by National Technology and Engineering Solutions of Sandia, LLC, a wholly owned subsidiary of Honeywell International, Inc., for the U.S. DOE’s National Nuclear Security Administration under contract DE-NA-0003525. 
\end{ack}

\printbibliography


\end{document}